\documentclass[jcp, floatfix, nobibnotes, reprint, superscriptaddress]{revtex4-1}
\usepackage{docs}%
\usepackage{siunitx}
\usepackage{amsmath}
\DeclareSIUnit[number-unit-product = {\,}]
\cal{cal}
\DeclareSIUnit\kcal{\kilo\cal}
\DeclareSIUnit[number-unit-product = {\,}]
\Btu{Btu}
\DeclareSIUnit[number-unit-product 
= {\,}]
\Fahr{\degree F}
\DeclareSIUnit[number-unit-product = {\,}]
\usepackage{bm}%
\usepackage[colorlinks=true,linkcolor=blue]{hyperref}%
\usepackage{graphicx}
\usepackage[utf8]{inputenc}
\usepackage{tabularx}
\usepackage[usenames,dvipsnames]{color}
\usepackage{dcolumn}
\usepackage{epstopdf}
\usepackage{afterpage}
\usepackage{setspace}
\usepackage{multirow}
\usepackage{dsfont}
\usepackage{sidecap}

\DeclareMathAlphabet\mathbfcal{OMS}{cmsy}{b}{n}

\setcitestyle{super}

\begin{document}

\title{Encrypted machine learning of molecular quantum properties}
\author{Jan Weinreich}

\affiliation{University of Vienna, Faculty of Physics, Kolingasse 14-16, AT-1090 Wien, Austria}
\affiliation{University of Vienna, Vienna Doctoral School in Physics, Boltzmanngasse 5, 1090 Vienna, Austria}
\author{Guido Falk von Rudorff}
\affiliation{University Kassel,
Department of Chemistry, 
Heinrich-Plett-Str.40,34132 Kassel, Germany}
\author{O. Anatole von Lilienfeld}
\email{anatole.vonlilienfeld@utoronto.ca}
\affiliation{Vector Institute for Artificial Intelligence, Toronto, ON, M5S 1M1, Canada}
\affiliation{Departments of Chemistry, Materials Science and Engineering, and Physics, University of Toronto, St. George Campus, Toronto, ON, Canada}
\affiliation{Machine Learning Group, Technische Universit\"at Berlin and Institute for the Foundations of Learning and Data, 10587 Berlin, Germany}

\date{\today}

\begin{abstract}
Large machine learning models with improved predictions have become widely available in the chemical sciences. 
Unfortunately, these models do not protect the privacy necessary within commercial settings, prohibiting the use of potentially extremely valuable data by others. Encrypting the prediction process can solve this problem by double-blind model evaluation and prohibits the extraction of training or query data.
However, contemporary ML models based on fully homomorphic encryption or federated learning are either too expensive for practical use or have to trade higher speed for weaker security. 
We have implemented secure and computationally feasible encrypted machine learning models using oblivious transfer enabling and secure predictions of molecular quantum properties across chemical compound space. However, we find that encrypted predictions using kernel ridge regression models are a million times more expensive than without encryption. This demonstrates a dire need for a compact machine learning model architecture, including molecular representation and kernel matrix size, that minimizes model evaluation costs.
\end{abstract}

\maketitle 

\bigskip

\section{Introduction}
\begin{figure}[h!]
          \centering           
          \includegraphics[width=0.8\columnwidth]{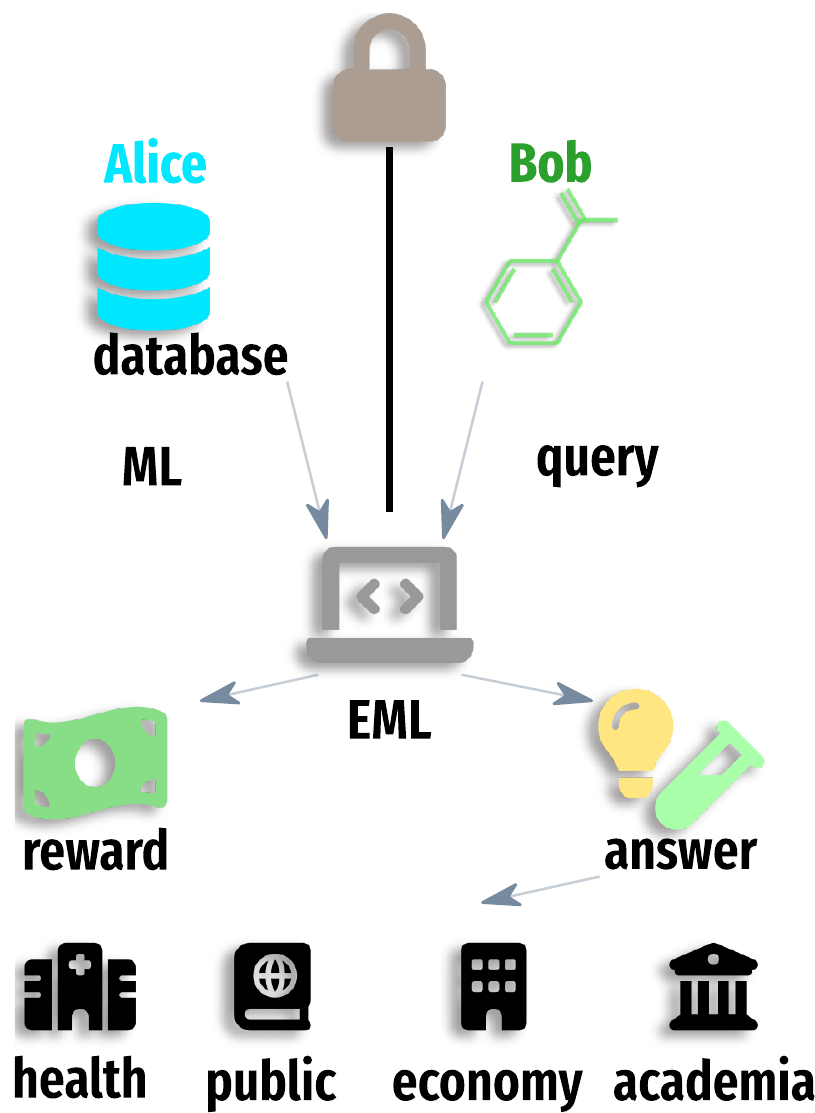}  
          \caption{
          Bob the user \textcolor{LimeGreen}{\textbf{B}}, working in health, company or academic or public sector, wants to predict properties of query molecules:
          First Alice \textcolor{SkyBlue}{\textbf{A}}, holder of a large database, and \textcolor{LimeGreen}{\textbf{B}} agree on a molecular representation and the EML protocol. Next \textcolor{SkyBlue}{\textbf{A}} trains EML on her local node with unencrypted calculations. Subsequently, both parties jointly compute the query prediction by exchanging encrypted information without disclosure.
          Finally, the prediction is revealed to \textcolor{LimeGreen}{\textbf{B}} while \textcolor{SkyBlue}{\textbf{A}} will never see the query data.
     \textcolor{LimeGreen}{\textbf{B}} \textit{vice versa} cannot extract any information about the hidden reference data owned by \textcolor{SkyBlue}{\textbf{A}}. All protocol steps are privacy-preserving and both parties effectively form an encrypted double-blind ML oracle.
          }
     \label{fig:concept} 
 \end{figure}
 
The global amount of information has grown exponentially over time. Seagate, a large data storage company, projects it to reach 181 zettabytes by 2025\footnote{Source: statista \url{https://www.statista.com/topics/1464/big-data/#topicHeader__wrapper}, \url{https://www.seagate.com/files/www-content/our-story/trends/files/idc-seagate-dataage-whitepaper.pdf}}. 
Countless machine learning (ML) applications are based on this wealth of data, as reflected in a rapidly growing number of ML publications\cite{mlperyear}. 
Still, especially sensitive data is not publicly accessible, preventing innovative ML innovations in these fields. The main issue is that the evaluation of ML models is not double-blind: 
The user querying the ML model can gather information about the training set and discloses all information about the query.
The holder of the database can accumulate huge amounts of querying user data posing a threat if the server is under attack from a third party. 
This is especially relevant considering the fast-growing use of cloud computing\footnote{Source: statista \url{https://www.statista.com/topics/1464/big-data/#topicHeader__wrapper}, \url{https://www.seagate.com/files/www-content/our-story/trends/files/idc-seagate-dataage-whitepaper.pdf}} and number of cyber attacks. End-to-end encryption cannot solve this issue as data is usually processed in unencrypted form. 

Decisions based on knowledge derived from protected data without revealing any data would warrant immediate benefits: Potentially relevant fields include modeling health data or sharing predictions evaluated on protected databases. Furthermore, double-blind ML evaluation may reduce customers' hesitations to send sensitive medical data to the cloud, allowing, for instance, more personalized health recommendations -- without giving away private information. 
From the viewpoint of chemistry and medical sciences, a potential application is commercial data from pharmaceutical companies, since a considerable amount is invested in various screening approaches (\textit{in vivo} and \textit{in vitro}).
While the collected data sets are relevant for developing new pharmaceuticals, they are generally not published. 
Currently, nondisclosure agreements are the only way for the chemical industry to provide academia with protected data. However, this comes with legal and economic risks as well as bureaucratic barriers. 
To give an idea of the value of privacy: A ballpark estimate of post-approval R\&D costs of a new drug ranges between \$1.8 to \$2.8 billion where much of these costs are clinical trials\cite{DIMASI201620, Paul2010HowTI,doi:10.1056/NEJMp1500848}.
In particular, toxicological assays are critical for the development of drug candidates\cite{toxy,knowTox}, new nanomaterials\cite{doi:10.1021/es802388s}, or pesticides, but can take many years and millions of dollars\cite{https://doi.org/10.1111/risa.13810, krewski} to complete. These time and cost constraints are a substantial bottleneck for the innovation of new substances. 
\\
An additional aspect is that the emergence of ML has made modern research more dependent on access to high-quality datasets. 
Compiling new impactful datasets is only possible with the required funding to perform lab experiments. Public data often originates from several different sources, resulting in inconsistencies\cite{meltingpointsarehorrible}. 
Access to predictions based on secret but single-origin high-quality measurements could help mitigate these problems. 
\\
Driven by this vision, a growing number of institutions is considering using ML models that allow hiding training data, as can be seen in recent projects such as the \textit{melloddy} initiative\footnote{\url{https://www.melloddy.eu/y2announcement}}. 
Multiple computational approaches for privacy-preserving ML have been developed; a popular example is federated learning\cite{adnan2022federated, pmlr-v54-mcmahan17a, fedjax2020github, hard2019federated, choquettechoo2021capc,SAV2022100487, Aggarwal2021FedFaceCL} where several data sets from different data holders are used to compute a local gradient for subsequent updates of a global model. In particular, Zhu et al\cite{white_federated} have implemented federated learning for molecular properties. Despite many advantages, 
if not properly addressed, federated learning can show several security risks. This is particularly the case when participants are allowed to deviate from the predefined machine learning protocol (in a \textit{malicious} adversary setting).
When training a federated learning model, each potentially \textit{malicious} participant can send false data on purpose\cite{shumailov2021manipulating} to prevent learning of the global model\cite{https://doi.org/10.48550/arxiv.2201.12675,https://doi.org/10.48550/arxiv.2202.00580}\footnote{possible risks of the mellody platform\url{https://www.melloddy.eu/blog/it-security-of-the-melloddy-platform}}. Furthermore, in an iterative procedure, any participant could compare the last global model with the previous state. This allows probing where the update of other data holders had the greatest impact to detect points that likely exist in the other data sets. In certain scenarios federated learning models allow unnoticed extraction of training data\cite{fowl2022robbing}. 

In this study we achieve double-blind ML prediction of molecular quantum properties by competitive cooperation, a.k.a. \textit{coopetition}: Two competitive parties that do not trust each other cooperate in exchanging encrypted pieces of information to evaluate the ML model as illustrated in Fig.~\ref{fig:concept}. Our encrypted ML (EML) protocol ensures that the data holder maintains access control to the model at all times. More specifically, we have considered a two-party setting with Alice \textcolor{SkyBlue}{\textbf{A}} the data holder and Bob \textcolor{LimeGreen}{\textbf{B}} the user querying the machine learning model. We will keep this color highlighting consistently throughout the following.
Neither \textcolor{SkyBlue}{\textbf{A}} nor \textcolor{LimeGreen}{\textbf{B}} reveal private data when querying the oracle.

To summarize the key properties and the threat model of the algorithm:
No central server is needed since oblivious transfer removes the need for a central entity as required in federated learning. Only the party that owns the data has access to weights, and we do not provide variance in predictions because we only use a single model split. The algorithm is based on encryption and is safe against a dishonest majority\cite{mascot}. The amount of data that can be recovered from a single prediction depends on what an adversary already knows about the individual whose privacy is at risk. Reconstruction attacks on training data have had limited success and researchers have focused mainly on membership inference\cite{membership_attack}, which can be used as a basis for reconstruction attacks. It is also possible to extract memorized private information from deployed language models\cite{languagemodelrec}. Regarding our approach, we assume that there is a secure communication channel between the two parties. Given the security of the oblivious transfer protocol, the data owner cannot learn anything about the query. 
While we have not found an example of such an attack in the literature the querying party, might send designed queries in an attempt to reconstruct the decision boundary of the kernel-ridge regression algorithm based only on the predicted values. However, we cannot rule out that it is possible to construct an attack in this manner.

Our solution to double-blind evaluation consists of an encrypted ML oracle based on kernel ridge regression\cite{Vapnik1998, anatolebook} and gives a single scalar value per query. Testing encrypted predictions for molecular properties reveals that the results of unencrypted calculations are exactly reproduced. We find that compact ML representations are superior in terms of cost per prediction and show higher numerical stability. 
 
\section{Methods}
 
\subsection{Oblivious transfer versus fully homomorphic encryption}

\begin{figure}[htb]
          \centering           
          \includegraphics[width=0.25\textwidth]{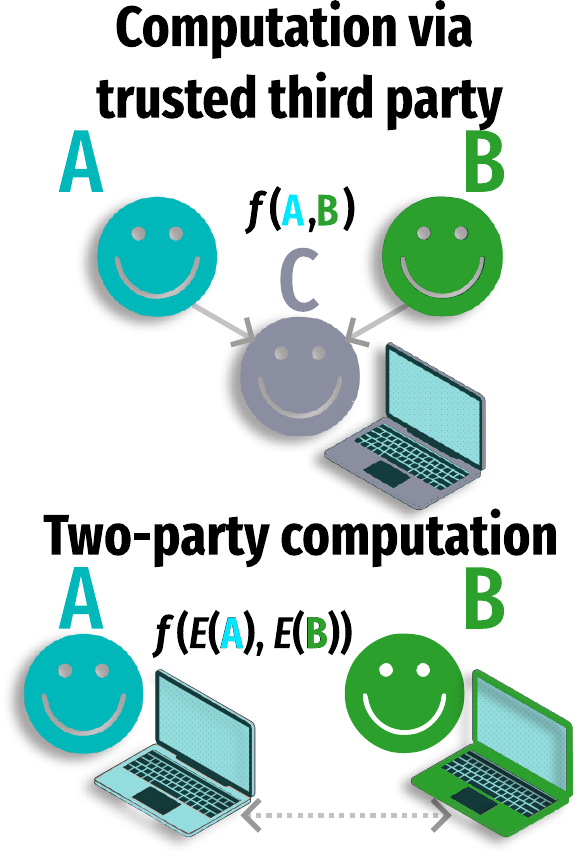}
          \caption{
          Parties Alice and Bob with private data 
          \textcolor{SkyBlue}{\textbf{A}} and \textcolor{LimeGreen}{\textbf{B}} respectively compute the value of a publicly known function $f$ that depends on their private inputs.
          Two scenarios are considered: Alice and Bob submit all their data to a trusted third party Carol \textcolor{Gray}{\textbf{C}} who subsequently performs the required calculations. 
          In the second case, two-party computation\cite{Yao1982ProtocolsFS,Yao1986HowTG} allows removing \textcolor{Gray}{\textbf{C}}. Instead, Alice and Bob exchange encrypted $E$ information packages via oblivious transfer\cite{10.1145/62212.62215} to evaluate the function $f$.
          }
     \label{fig:sketch_MPC}
 \end{figure}

Ideas of encrypted calculations of arbitrary functions were first hypothesized in the late 70s\cite{Rivest1978}. The archetypal problem solved in this context was Yao's Millionaires' problem: Two wealthy individuals with money amount $x_1$ and $x_2$ want to know if $x_1> x_2$ is true or false without revealing exact amounts\cite{Yao1982ProtocolsFS, Yao1986HowTG} $x_1$ and $x_2$. 
\\
However, the extension allowing encryption of any calculation took until 2009 when the first algorithm for a fully homomorphic encryption scheme was described. This allows fully encrypted addition and multiplication of any number and as such evaluation of any real function $f$.
To explain what is meant by computation on encrypted data, we illustrate the addition of numbers $x_3=x_1+x_2$. The addition is performed with encrypted $E$ representations or ciphertexts $c_1$ and $c_2$ of numbers with a public key $p_k$. The first ciphertext is $c_1 = E(p_k , x_1)$ and the second $c_2 = E(p_k , x_2)$. The decryption $\text{D}$ of the addition using the secret key $s_k$ results in the correct number as follows,
\begin{align}
    x_3 &=\text{D}(s_{k}, c_{3}) = 
    \text{D}(s_{k}, c_{1} + c_{2}) \\
    &= \text{D}(s_{k}, c_{1}) + \text{D}(s_{k}, c_{2}) = x_{1} + x_{2}~.
\end{align}
Such encrypted calculations are called fully homomorphic encryption\cite{10.1145/1536414.1536440},
\textit{fully} because any function can be evaluated, and \textit{homomorphic} meaning \textit{same shape} because fully homomorphic encryption conserves relations between numbers in the encrypted space. A benefit of fully homomorphic encryption is that it does not require communication between the parties that own the private data. Computations are performed \textit{offline}. This may also be viewed as a disadvantage since parties cannot query discovery-based requests where ad hoc access to results is necessary. A downside of fully homomorphic encryption is that computations are quite expensive. Furthermore, a central server is needed to perform the calculations only after receiving all encrypted information at once.
\\
An alternative method for privacy-preserving function evaluation is multi-party computation\cite{Yao1982ProtocolsFS,Yao1986HowTG,cryptoeprint:2020:521}. In the case of two parties, multi-party computation reduces to two-party computation (s. Fig.~\ref{fig:sketch_MPC}). To perform an encrypted calculation with public function $f$ parties Alice and Bob exchange encrypted chunks of data without disclosing anything about their private inputs \textcolor{LimeGreen}{\textbf{B}} and \textcolor{SkyBlue}{\textbf{A}}. The exchange of data packages is conducted via oblivious transfer\cite{10.1145/62212.62215}. An oblivious transfer protocol consists of at least one sender and a receiver. The sender sends an \textit{oblivious} amount of information packages to the receiver i.e. much more information than necessary for each round of communication. To the sender, it remains \textit{oblivious} which bit of information was obtained by the receiver. During the oblivious transfer evaluation of the function, the roles of the receiver and sender are frequently interchanged, but at no point has any party enough information to reconstruct intermediate results. Remarkably, this rather unintuitive way of exchanging information allows encrypted evaluation of any real function\cite{Schoenmakers2011}.
The key advantage of two-party computation via oblivious transfer and in particular of the protocol called \underline{m}alicious \underline{a}rithmetic \underline{s}ecure \underline{c}omputation with oblivious transfer\cite{mascot} (MASCOT) is the small computational cost compared to fully homomorphic encryption and other implementations of multi-party computation. MASCOT provides security against a dishonest majority of attackers with \textit{malicious} intent. As for all multi-party computation algorithms, continuous communication between all involved parties is needed, so the transfer of data is the main computing bottleneck. In MASCOT, floating-point numbers are translated into a finite integer representation. To avoid overflow errors the numerical precision $\mathcal{P}$ (s. detailed explanation of $\mathcal{P}$ in SI.~sec.~C) can be increased to allow representing larger numbers with better resolution.

\subsection{Encrypted machine learning of molecular properties}

\begin{figure}[htb]
          \centering           
\includegraphics[width=0.4\textwidth]{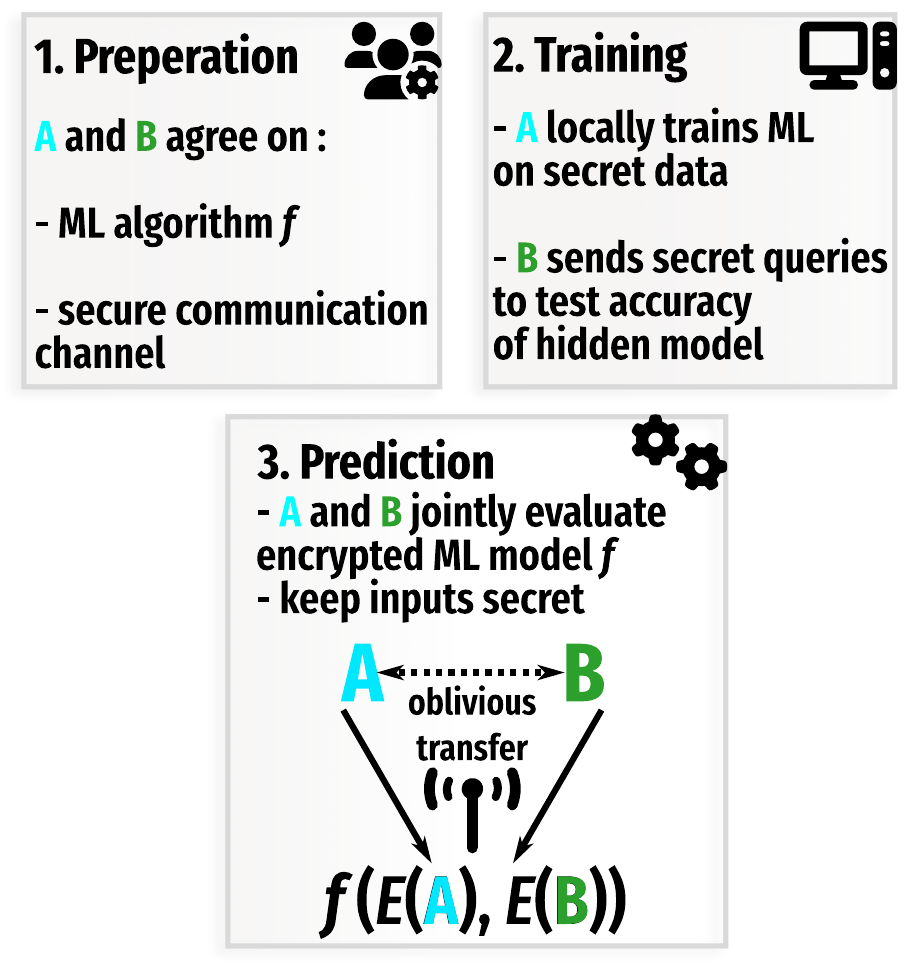}
          \caption{
            Three-stage process of encrypted machine learning (EML) predictions for two parties, the data holder \textcolor{SkyBlue}{\textbf{A}} providing training data and \textcolor{LimeGreen}{\textbf{B}} the user submitting the query: Preparation (1), training (2) and evaluation (3). The dashed arrow indicates the encrypted exchange via oblivious transfer for evaluation of $f$ on the private inputs.
          }
     \label{fig:concept2} 
 \end{figure}

\subsubsection{Encrypted kernel ridge regression}

Alice \textcolor{SkyBlue}{\textbf{A}} holds secret training data and collaborates with Bob \textcolor{LimeGreen}{\textbf{B}} the user by providing encrypted ML (EML) predictions to his queries. \textcolor{LimeGreen}{\textbf{B}} should not be able to learn anything about the training set, \textcolor{SkyBlue}{\textbf{A}} should not learn anything about the query of \textcolor{LimeGreen}{\textbf{B}}. Only the prediction is sent to \textcolor{LimeGreen}{\textbf{B}} while the calculations cannot be inspected or manipulated by either party.
We address this problem by encrypting the ML predictions using the MASCOT protocol discussed in the previous section. All following mathematical expressions are colored according to access to the respective data before, during, or after the encrypted prediction.
\\
Setting up the ML oracle can be separated into three steps shown in Fig.~\ref{fig:concept2}: First, both parties agree on the same mathematical form to represent the data.  In the case of movie preferences, this could be a vector that contains location and age. For cloud-based services, it could be private data such as heart pressure, blood sugar, or pulse. 
For secret new drug-like molecules, we use molecular representation vectors such as the Coulomb matrix\cite{coulomb} (CM), or the FCHL19\cite{FCHL, felixFCHL} that require three-dimensional nuclear coordinates and charges. Note that FCHL19 is a local representation that allows one to compare atomic environments between different molecules with each other. However for demonstration purposes and because it allows direct timing benchmark comparisons we will treat FCHL19 as a flattened global representation vector like the CM.
\\
Secondly, both parties agree on an ML protocol $f$, here kernel ridge regression\cite{Vapnik1998, anatolebook}. Kernel ridge regression is a supervised learning method in which for each prediction the features of the query instance are compared against all training instances and weighted by regression coefficients. In the next phase, \textcolor{SkyBlue}{\textbf{A}} trains a hidden ML model on her local machine. \textcolor{SkyBlue}{\textbf{A}} locally computes the input representation vectors $\textcolor{SkyBlue}{\mathbf{X}_i}$ that can correspond to any set of labels that show good correlation with the quantity of interest $y$. The kernel ridge regression weights $\textcolor{SkyBlue}{\boldsymbol{\alpha}}$ are obtained by solving a system of equations,
\begin{align}
    \textcolor{SkyBlue}{\boldsymbol{\alpha}} = \left[\textcolor{SkyBlue}{\mathbf{K}} + \textcolor{SkyBlue}{\lambda} \cdot \mathds{I}\right]^{-1} \textcolor{SkyBlue}{\boldsymbol{y}}.
\end{align}
All quantities in the upper equation are known to \textcolor{SkyBlue}{\textbf{A}} notably the values $\textcolor{SkyBlue}{\boldsymbol{y}}$ of hidden data. The elements of the training kernel matrix $\textcolor{SkyBlue}{\mathbf{K}}$ are computed with Gaussian functions,
\begin{align}
    (\mathbf{K})_{i,j} = \exp{\left[ -\frac{\vert \vert \mathbf{X}_i - \mathbf{X}_j \vert \vert^{2}_{2}}{2 \sigma^2} \right]}~,
    \label{eq:kernel}
\end{align}
where the elements $i,j$ are contained in the hidden training set and $||.||_2$ denotes the euclidean norm. The hyperparameter $\sigma$ is shared with \textcolor{LimeGreen}{\textbf{B}} while $\textcolor{SkyBlue}{\lambda}$ is kept private. Next, \textcolor{LimeGreen}{\textbf{B}} calculates the representation vector $\textcolor{LimeGreen}{\mathbf{X}_Q}$ of the query entities on his local machine. Next, the training weights $\text{E}( \textcolor{SkyBlue}{{\boldsymbol{\alpha}}} )$, input representation vectors $ \text{E}(\textcolor{SkyBlue}{{\mathbf{X}}_i})$ and the query representations $\text{E}( \textcolor{LimeGreen}{{\mathbf{X}}_Q})$ are encrypted (recall that $E$ is encryption and $D$ decryption). This process takes place during a prepossess phase after establishing a secure communication channel between the two parties. In practice, we perform all calculations using a virtual network on a single machine.
\\
Subsequently, the following expression for encrypted kernel ridge regression prediction is evaluated,
\begin{equation}
\begin{split}
    f &= D ( \textcolor{LimeGreen}{\text{E}(y) [E({\mathbf{X}}_{Q})]} \\ &= 
    D \left(
    \sum_{i}^{\textcolor{SkyBlue}{N}} E(\textcolor{SkyBlue}{{\alpha}_i})~ E(k)[\textcolor{SkyBlue}{E({\mathbf{X}}_{i})},\textcolor{LimeGreen}{E({\mathbf{X}}_{Q}}); \sigma]
    \right)
    ~.
    \label{eq:crypto_mapping}
\end{split}
\end{equation}
It is essential that the kernel values $E({k})[\textcolor{SkyBlue}{{E(\mathbf{X}}_{i}}),E(\textcolor{LimeGreen}{{\mathbf{X}}_{Q}})]$ are not known to either party. Otherwise, participants could probe kernel elements ${k}$ by repeatedly querying the oracle to obtain the compound space spanned by the training molecules. For the same reason, the distances ${d}_{iQ}=\vert \vert \textcolor{SkyBlue}{{\mathbf{X}}_{i}} - \textcolor{LimeGreen}{{\mathbf{X}}_{Q}} \vert \vert_{2}^{2}$ between the training set and the query molecule\cite{fenner2020privacypreserving} are never disclosed. Next \textcolor{LimeGreen}{\textbf{B}} may evaluate a few encrypted samples to validate the consistency of the hidden predictions. If found to be necessary, \textcolor{SkyBlue}{\textbf{A}} can increase the training set size or data diversity in hope of improving the accuracy of the model. In the prediction phase, $f$ is evaluated via oblivious transfer without disclosing $E(\textcolor{SkyBlue}{{\alpha}_i}), E(\textcolor{SkyBlue}{{\mathbf{X}}_{i}}),E(\textcolor{LimeGreen}{{\mathbf{X}}_{Q}})$. Finally, the decrypted plaintext predictions are send to \textcolor{LimeGreen}{\textbf{B}} while \textcolor{SkyBlue}{\textbf{A}} could obtain a reward in form of a payment for the prediction provided. Effectively, both parties are part of an ML oracle with a true black-box character. 

We use learning curves to quantify the error of the predictions w.r.t. the reference values measured as the mean absolute error (MAE) as a function of the size of the training set $N$. Learning curves are helpful to understand the efficiency of ML models and are generally found\cite{Vapnik1998} to be linear on a log-log scale,
\begin{align}
    \log{ \left( \frac{ \text{MAE} }{\text{unit}} \right) }   \approx I -  S \cdot \log{(N)}~,
\end{align}
where $I$ is the initial error and $S$ is the slope indicating the improvement of the model given more training data.

\section{Results and discussion}
\label{sec:results}

\subsection{Encrypted Kernel predictions: \textit{malicious} security for computational chemistry}

Next, we demonstrate encrypted ML predictions for fictitiously confidential chemical data. Predicting the stability of molecules is a key problem in computational chemistry and is well described by solving the Schr{\"o}dinger equation and atomization energies, the energy contained in all bonds of a molecule. However, solving the Schr{\"o}dinger equation comes at high computational costs: For instance, costs for solutions of a density functional theory calculation scale to the cubed power with the number of atoms. To give a very rough estimate, computing a molecular dataset with $\sim20000$ molecules of the size of aspirin with coupled cluster singles and doubles\cite{CISD} scaling with the seventh power of system size would consume 20000 CPU hours\cite{Heinen_2020} -- even for a relatively small basis set such as def2-SVP\cite{doi:10.1063/1.467146}.
\begin{figure}[htb]
          \centering           
     \includegraphics[width=0.35\textwidth]{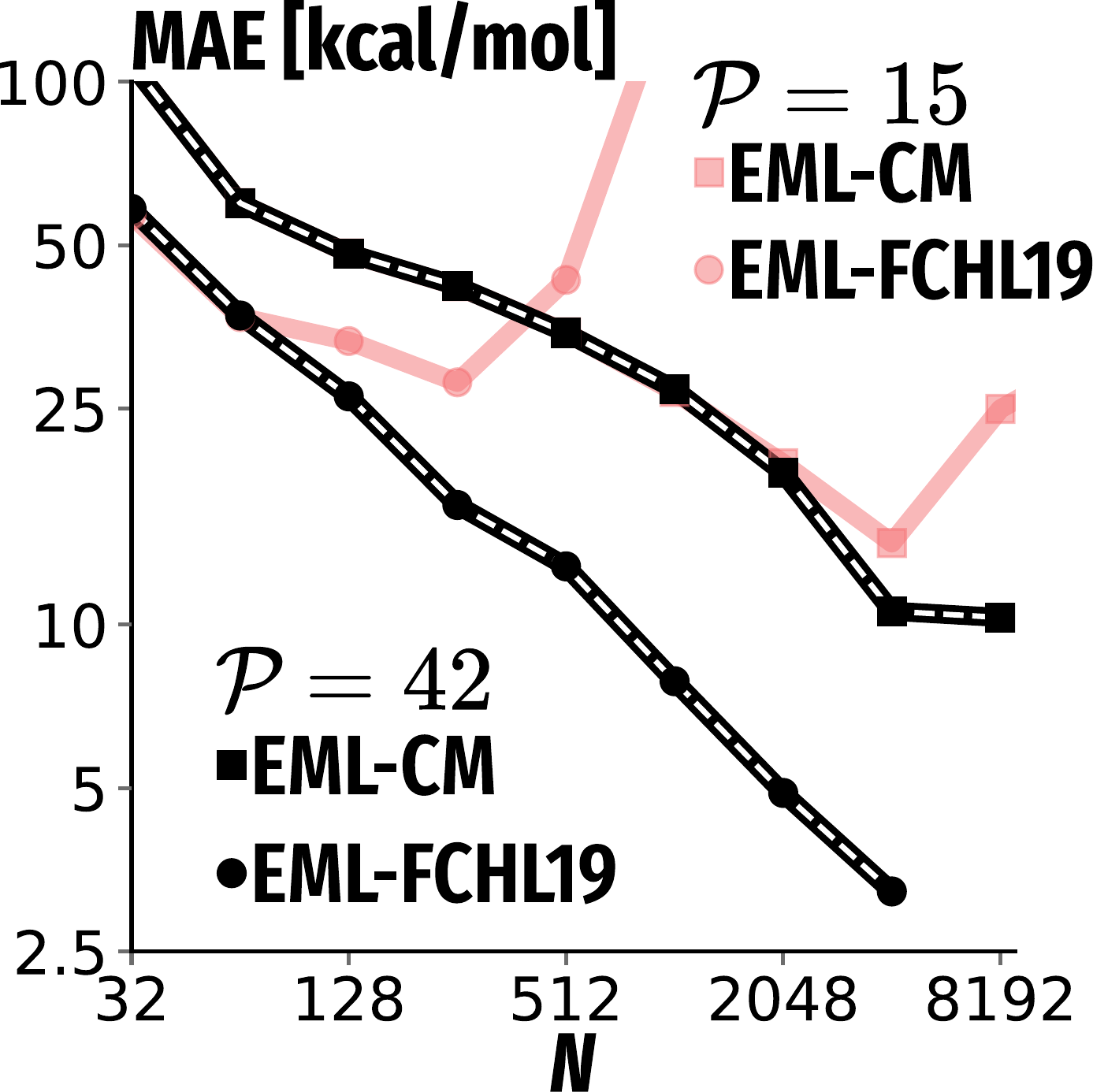}
          \caption{
            Prediction error for a random subset of QM9\cite{qm9} atomization energies as a function of training set size $N$: Mean absolute error (MAE) from encrypted machine learning (EML). Results are shown for two different molecular representations, the coulomb matrix\cite{coulomb} (CM) and FCHL19\cite{FCHL, felixFCHL}. The numerical precision of fixed-point number representation of floating point numbers is set to either $\mathcal{P}=15$ or $42$. The dashed white lines superimposed with the $\mathcal{P}=42$ EML results, show the learning curves computed with unencrypted python predictions for comparison.
          }
     \label{fig:eml_1st_lrncrvA} 
 \end{figure} 
 
\begin{figure}[htb]
          \centering           
\includegraphics[width=0.4\textwidth]{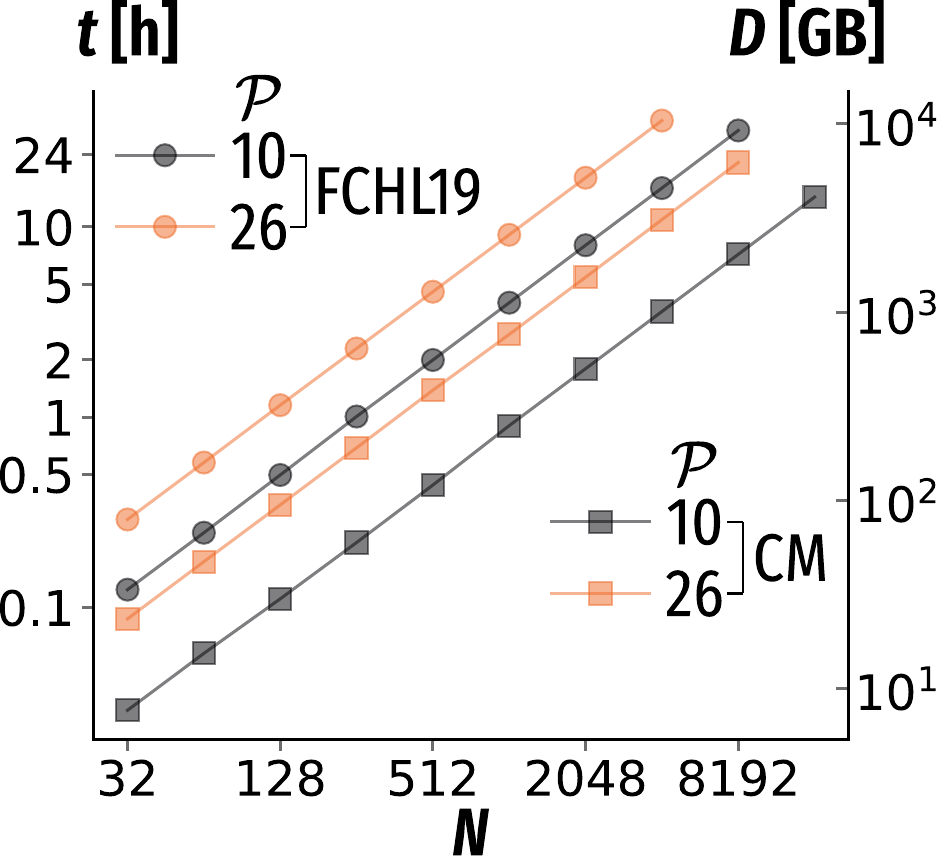}
          \caption{
           Encrypted machine learning (EML) prediction times averaged per molecule (left) and transferred data $D$ (right) for Coulomb matrix\cite{coulomb} (CM) and FCHL19\cite{FCHL, felixFCHL} representations. Results for two different numerical accuracy settings $\mathcal{P}=10, 26$ are shown.
          }
     \label{fig:eml_1st_lrncrvB} 
 \end{figure} 
Such high computational costs underline the value of high-level computational data. As a potential scenario, we consider company \textcolor{SkyBlue}{\textbf{A}} providing encrypted ML predictions and an industrial customer \textcolor{LimeGreen}{\textbf{B}} with interest in 20 secret molecules but without time, experience or access to software to perform calculations. We have computed learning curves of atomization energies using encrypted predictions with the QM9 database\cite{qm9} of organic molecules with up to $N=8192$ compounds. The resulting learning curves using the CM\cite{coulomb} and FCHL19\cite{FCHL, felixFCHL} representations are shown in Fig.~\ref{fig:eml_1st_lrncrvA}. The deviation from the unencrypted case only amounts to numerical noise and cannot be identified visually in the learning curves. Hence, we find that EML accurately reproduces unencrypted predictions. As expected, time $t$, as well as data traffic $D$ per prediction, increases linearly with the number of training points $N$ (s. Fig.~\ref{fig:eml_1st_lrncrvB}). Furthermore, there is a striking difference between the FCHL19\cite{FCHL, felixFCHL} representation with $L = 18720$ entries that takes more than twice as long (1 hour at $N=128$) for a single prediction than the CM\cite{coulomb} ($L=351$). Contrary to FCHL19 the CM representation contains less information i.e. no angles or local environments resulting in a larger MAE. Overall, data transfer $D$ between parties is the main computational bottleneck for prediction\cite{mascot} explaining the near-perfect correlation between $D$ and $t$ (s. Fig.~\ref{fig:eml_1st_lrncrvB}).
\\
Consequently, compact representations such as the CM reduce the prediction time by reducing $D$. The role of compact representations becomes more evident when studying QM9 learning curves (s. Fig.~\ref{fig:eml_1st_lrncrvA}) for lower numerical precision settings corresponding to faster predictions. For high numerical precision ($\mathcal{P}=42$) there is hardly any visible difference between the EML and kernel ridge regression learning curves (as in Fig.~\ref{fig:eml_1st_lrncrvA}). At $\mathcal{P}=15$, we find that the FCHL19 EML learning curve shows a dramatic deterioration for $N\geq128$ while the CM learning curve only begins to deviate at $N > 2000$. Although compact representations include less chemical information, they allow for larger training set sizes, given the same target accuracy as well as high numerical stability. If using representation vectors such as FCHL19 cannot be avoided because predictions with high accuracy w.r.t. the test set is needed the numerical precision $\mathcal{P}$ can be increased to avoid numerical instabilities. Fortunately,
there exists an optimal $\mathcal{P}$ with minimal computational cost and sufficient numerical precision. This is because $t$ increases only quadratically with $\mathcal{P}$, while the numerical deviation decays exponentially (SI Fig.~3d).

\subsection{Limitations and attack scenarios}
\label{sec:limitationsANDattacks}

The user \textcolor{LimeGreen}{\textbf{B}} can query the oracle with points for which reference values are known. A small error for the predicted values would suggest that similar points exist in the hidden training set. This attack will probably not be a threat, as it may require more points as contained in the hidden training set. On the other hand, this procedure can reassure \textcolor{LimeGreen}{\textbf{B}} that the hidden model provides reasonable predictions and that \textcolor{SkyBlue}{\textbf{A}} has not deliberately added incorrect training points. If \textcolor{LimeGreen}{\textbf{B}} knew the scaling rule of the kernel ridge regression ML oracle and the time needed per prediction \textcolor{LimeGreen}{\textbf{B}} might be able to guess the number of hidden training molecules. To address this issue \textcolor{SkyBlue}{\textbf{A}} could artificially increase the training set by adding a random number of duplicate training points. It is important to note that the ML oracle can only be trained and evaluated using a single training set split. Otherwise, the evaluation of ML models with different splits would leak the variance in addition to the predictions and which would enable attacks\cite{10.1145/2810103.2813677,reconstructing-training-data}.

An inherent problem of neural networks trained with hidden data is that the loss function gradient vanishes for training set points. In addition, generative adversarial networks are used to reverse engineer points in the training set\cite{haim2022reconstructing,fowl2022robbing,ctx28583922150005899}. Although our approach guarantees safety, this comes with increased computational costs compared to unencrypted calculations. In turn, we find that \textit{honest-but-curios} neural network predictions are orders of magnitude faster since the prediction speed does not depend on the number of training points (s. SI sec.~IV). However, the neural network protocol we have considered in the SI may not be safe against malicious attacks. Our MASCOT implementation of kernel ridge regression was the exact opposite in these two regards: Evaluation is relatively slow but secure. We find that encrypting ML predictions is a trade-off between security and computational efficiency.

\section{Conclusion}
\label{sec:conclusion}

The main advantage of our protocol is its safety against attackers with \textit{malicious} intent, as it is impossible to extract any molecular information, either from training or query instances, solely by evaluating encrypted predictions. 

The protocol eliminates the need for a trusted third party or central server, as required by fully homomorphic encryption. Instead, it requires only a secure communication channel between the two parties. Since the protocol is online no transfer of all the encrypted data to a single server is needed, contrary to fully homomorphic encryption. 
This also allows live predictions for new query molecules. The latter aspect is important for Bayesian exploration of chemical space, e.g. in the context of self-driving laboratories\cite{doi:10.1021/acs.accounts.2c00220} that would require \textit{ad hoc} predictions. We demonstrated that encrypted predictions of molecular properties based on EML are possible cf. Fig.~\ref{fig:eml_1st_lrncrvA}.
EML can be adapted to various properties and chemistries with negligible adaptation of the encrypted kernel ridge regression protocol.
Since EML does not require molecular representations as input, it may also be applied to pharmaceutical and private data from healthcare or finance.
\\
Our implementation was only possible thanks to recent developments in multi-party computation protocols\cite{mascot}. We note that added security comes at substantial additional computational costs with data transfer being the main computational bottleneck. Consequently, the compactness of the ML model, in the case of EML the kernel and the molecular representation play a crucial role. More specifically, we have demonstrated that verbose molecular representation vectors such as FCHL19\cite{FCHL, felixFCHL} allow for more accurate predictions than the more compact Coulomb matrix\cite{coulomb}. 
\\
As a result, users have to trade off the cost, accuracy, and security of the protocol. Our ball-park estimates indicate that a single molecular EML prediction is a million times more expensive than kernel ridge regression implemented in python code (s. Fig.~\ref{fig:eml_1st_lrncrvB}). For instance, approximately 250 GB of network traffic is needed for a single prediction at a modest training set size of $N = 512$ using the extended-connectivity fingerprint\cite{doi:10.1021/ci100050t,ecfp} which is often used in cheminformatics. Since there is a growing interest in maintaining privacy in ML it can be expected that future implementation of oblivious transfer will become much more efficient.

A goal of encrypted predictions is to enable decisions based on hidden data as if the knowledge leading to their actions was obtained by inspecting the secret data. However, since the prediction is encrypted, it is impossible to explain the actions that are solely based on the predictions. This lack of transparency may be problematic, as the model could have biases that cannot be explained by users unable to inspect the training data. It is an open question how encrypted predictions can be rationalized without inspecting the training set. One possible approach might be to understand the general behavior of the encrypted model without access to the underlying data, providing insight into the factors influencing the system's predictions.

\section{Supplemental material}

See supplemental material (SI) for more information on the numerical stability of the EML protocol and how numerical noise can be mitigated (SI. Fig.~1, 2). We also measure the quantitative influence of the representation length on the prediction time (SI Fig.~4). In the SI we also discuss encrypted neural network predictions of quantum properties based on weaker security (s. SI Fig. 5, 6, 7, 8).

\section{Data and Code availability}
\label{sec:data_avail}
The EML code to the high-level interface to multi-party computation protocol MASCOT\cite{mascot, mp-spdz} is available in the public GitHub repository at \url{https://github.com/janweinreich/EML/}. The repository also contains scripts for preparing the molecular quantum data for use with the EML protocol and an example of a neural network implementation of molecular property predictions using oblivious transfer. It also contains a permanent link to the input data to reproduce learning curves and encrypted predictions.

\section{Acknowledgments}
O.A.v.L. has received funding from the European Research Council (ERC) under the European Union’s Horizon 2020 research and innovation program (grant agreement No. 772834).
J.W. acknowledges support from the Faculty of Physics and supervision by C. Dellago at the University of Vienna. 
J.W. also acknowledges support from J.G. Brandenburg, in particular for proofreading the manuscript. O.A.v.L. has received support as the Ed Clark Chair of Advanced Materials and as a CIFAR AI chair.

\subsection{Dataset}

\label{sec:datasets}

To demonstrate the protocol for molecular property prediction, we use 30000 random molecules of the QM9\cite{qm9} data set with a random split into training and a random test set of 20 molecules. We predict the atomization energies, which measure the total energy necessary to dissociate a molecular compound into individual atoms. Hyperparameters are optimized with five-fold cross-validation for the largest training set size using unencrypted calculations with the quantum machine learning code\cite{qmlcode}. 
\\
The encrypted oblivious transfer calculations were performed in a local network with Intel(R) Xeon(R) E5-2650 v4 @ 2.20GHz CPUs. The number values for the reported timings may differ depending on the hardware.

\newpage
\newpage

\section{Third party material}
In Fig.~1,2,3 we included icons and modified them with permission under the license \url{https://fontawesome.com/license}.
\section{Declaration of Conflicting Interests}
We have no conflicting interests to declare.

\newpage

\section{SUPPLEMENTAL INFORMATION}

\subsection{Numerical Error and Precision}
\label{sec:numacc}

\begin{figure}[htb]
          \centering           
          \includegraphics[width=0.4\textwidth]{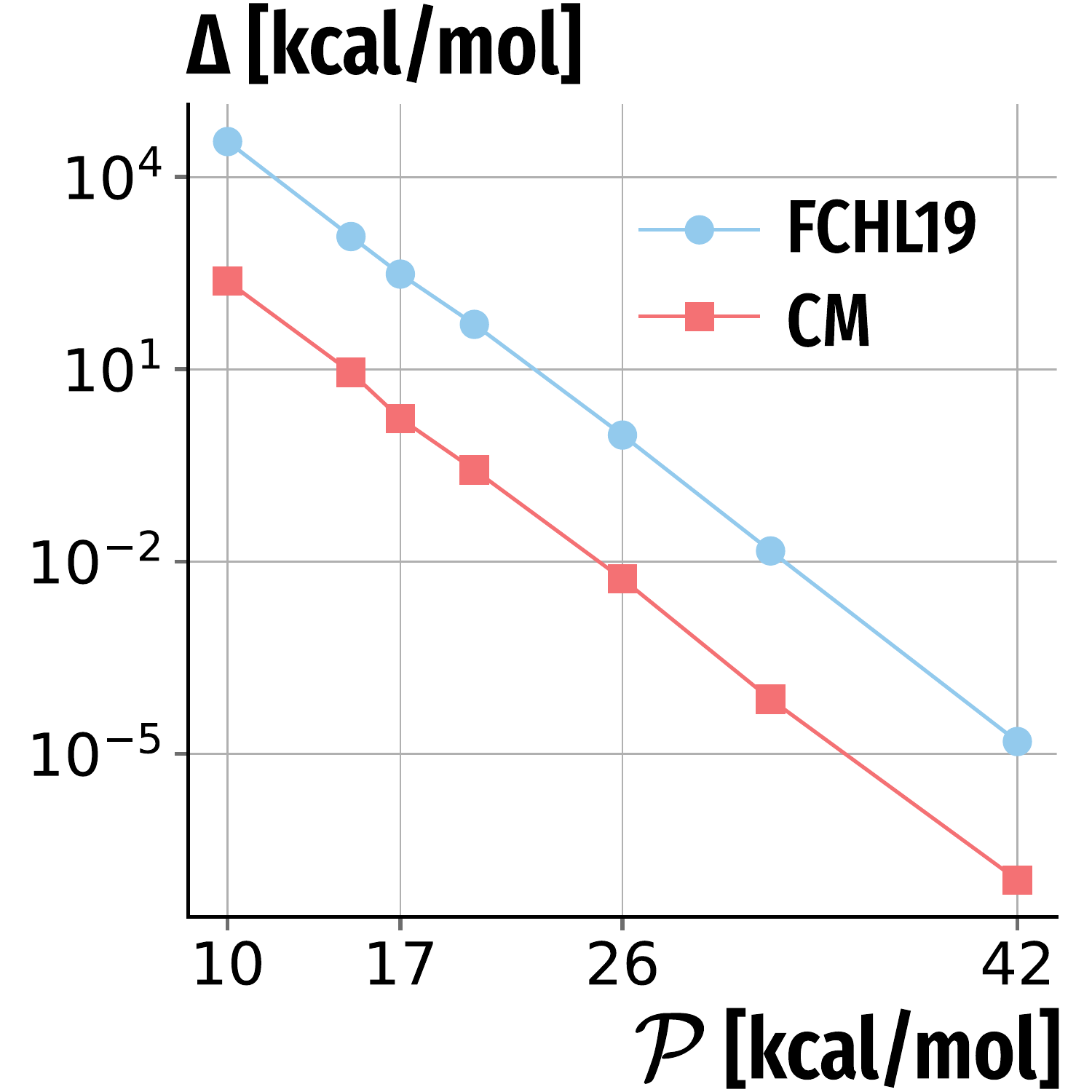}
          \caption{
          Average numerical error $\Delta$ between encrypted machine learning (EML) and plaintext python predictions for various accuracy settings $\mathcal{P}$ (c.f. Eq.~\ref{eq:numacc}) at training set size of $N=4097$. Results for two different molecular representations, the Coulomb Matrix\cite{coulomb} (CM) and FCHL19\cite{FCHL, felixFCHL}.
          }
     \label{fig:eml_accuracy} 
 \end{figure}

\begin{figure}[htb]
          \centering           
          \includegraphics[width=0.4\textwidth]{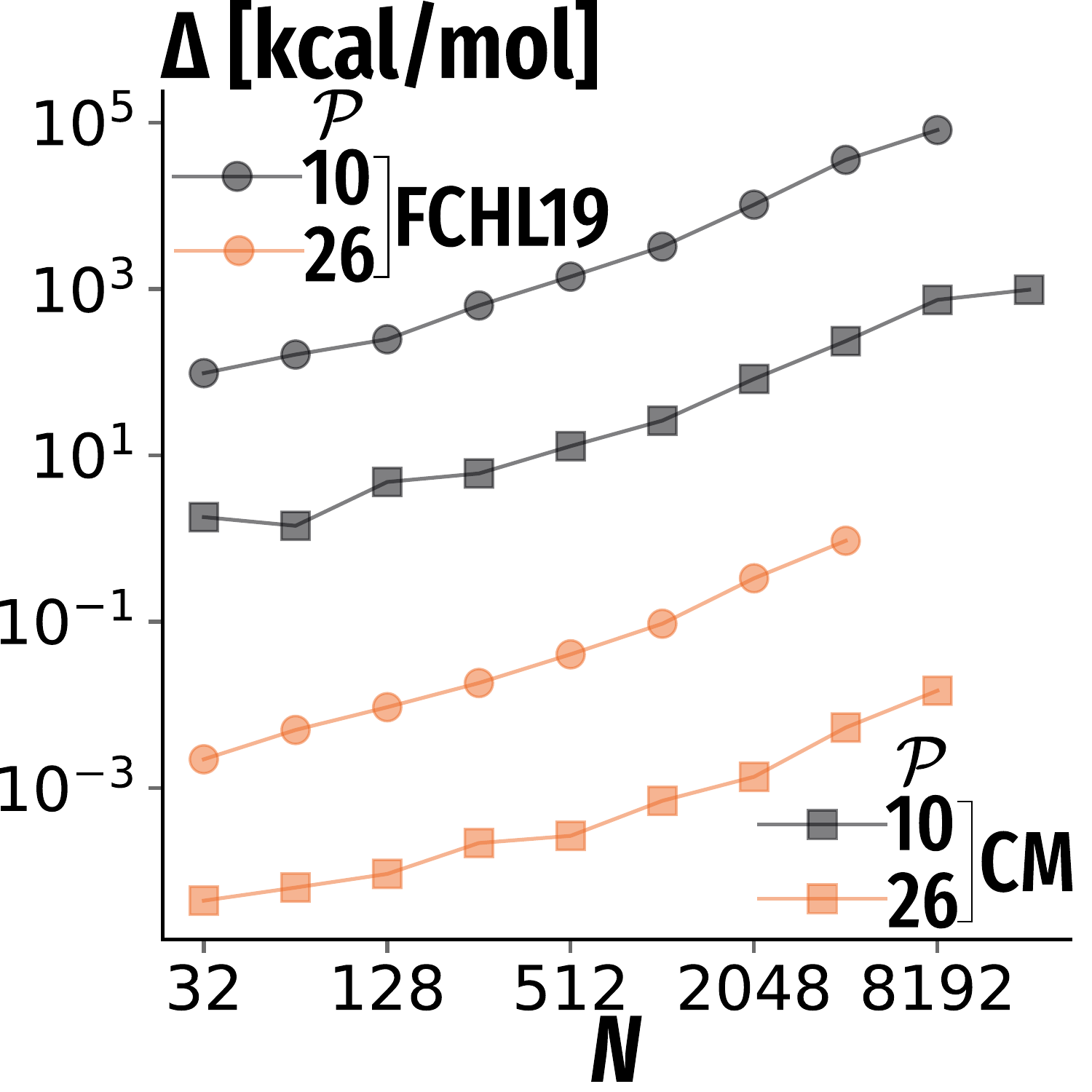}
          \caption{Average numerical deviation $\Delta$ between kernel ridge regression predictions with python and encrypted machine learning (EML) as a function of training set size $N$. Results for two different molecular representations, Coulomb Matrix\cite{coulomb} (CM) and FCHL19\cite{felixFCHL, FCHL}, and two different numerical accuracies $\mathcal{P}$ (c.f. Eq.~\ref{eq:numacc}) are shown.}
     \label{fig:eml_num_dev} 
 \end{figure}

\begin{figure}[htb]
          \centering           
          \includegraphics[width=0.4\textwidth]{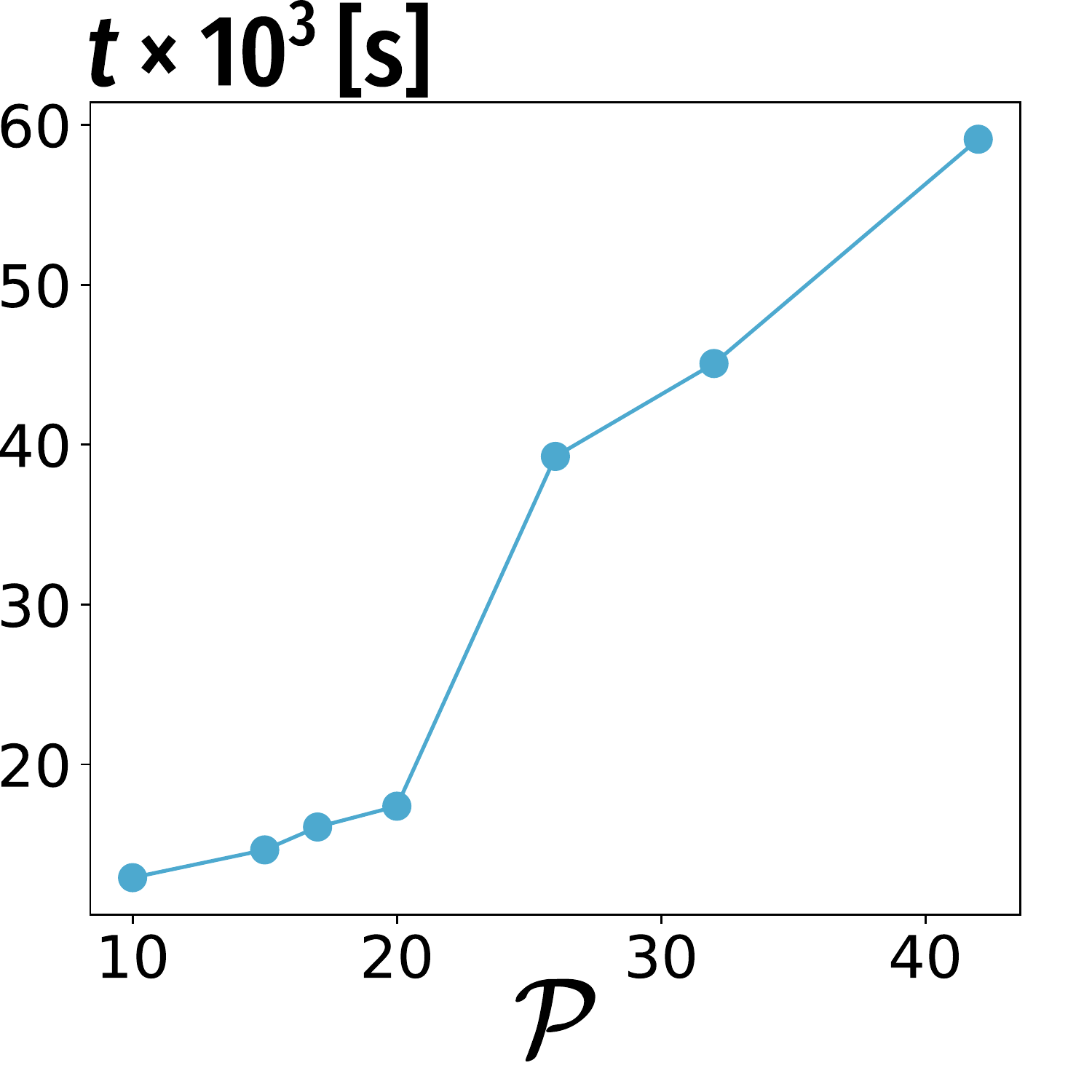}
          \caption{Per molecule prediction time $t$ for various numerical precision settings $\mathcal{P}$ using the Coulomb matrix at a training set size of $N=4096$.
          }
     \label{fig:TvsA} 
 \end{figure}

To control the numerical error the precision of the encrypted calculations can be controlled by the precision of the integer representation of floating-point numbers of the oblivious transfer protocol. The precision $\mathcal{P}$ is defined by two parameters\cite{mp_doc}: The first is the bit length of the decimal part $\mathcal{P}$. The second is the whole bit length of the fixed-point number $\mathcal{M}$. For simplicity, we summarize both in a single parameter by setting,
\begin{align}
\mathcal{P}:= \mathcal{M}-21 ~,
\label{eq:numacc}
\end{align}
as a single numerical precision parameter. Generally, the average error over all encrypted predictions $\widetilde{y}$ w.r.t. the test labels must be minimized.
Due to the additional numerical error of encrypted calculations, it must also be guaranteed that all predictions agree with python kernel ridge regression predictions.
To measure the numerical error only due to the encryption protocol we define the average numerical deviation as follows,
\begin{align}
   \Delta :=  \langle | \underbrace{ \widetilde{f}}_{\text{EML}}  - \underbrace{f}_{\text{python}} | \rangle ~.
     \label{eq:condition}
\end{align}
In Fig.~\ref{fig:eml_accuracy} we show the numerical error $\Delta$ as a function of numerical precision $\mathcal{P}$ and in Fig.~\ref{fig:eml_num_dev} for different training set sizes $N$ respectively. All other parameters are kept constant. We also show how the prediction time $t$ scales with the numerical precision $\mathcal{P}$ in Fig.~\ref{fig:TvsA}.

\subsection{Length of Input Features}

  \begin{figure}[htb]
          \centering           
          \includegraphics[width=0.4\textwidth]{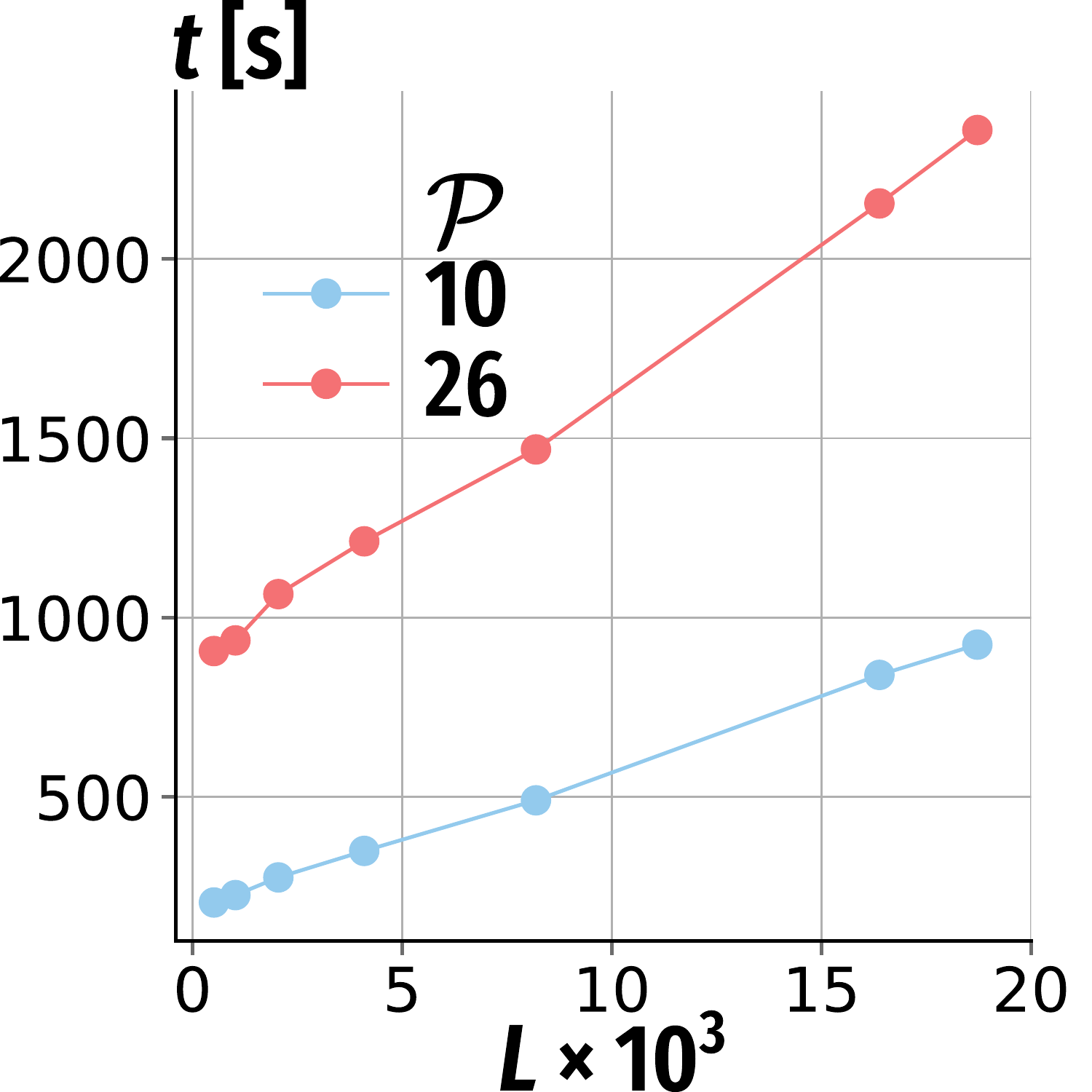}
          \caption{
            Encrypted machine learning (EML) prediction time $t$ and
          representation length $L$ using truncated FCHL19\cite{FCHL,felixFCHL} representation vectors for numerical accuracy levels of $\mathcal{P}=10, 26$ (s. Eq.~\ref{eq:numacc}) and $N=64$.
          }
     \label{fig:num_dev_len} 
 \end{figure}

We show the prediction time as a function of representation length $L$ (s. Fig.~\ref{fig:num_dev_len}) using truncated FCHL19\cite{FCHL,felixFCHL} molecular representation vectors. We find a linear scaling between $L$ and prediction time $t$. This is plausible given the element-wise distance evaluation of the euclidean norm. The numerical error $\Delta$, defined in Eq.~(\ref{eq:condition}), for different values of $L$ is shown in Fig.~\ref{fig:eml_num_dev}.

\subsection{Encrypted neural networks}

\subsubsection{Encrypted predictions}

\label{sec:enc_neural_networks}

  \begin{figure*}[htb]
          \centering           
    \includegraphics[width=0.6\textwidth]{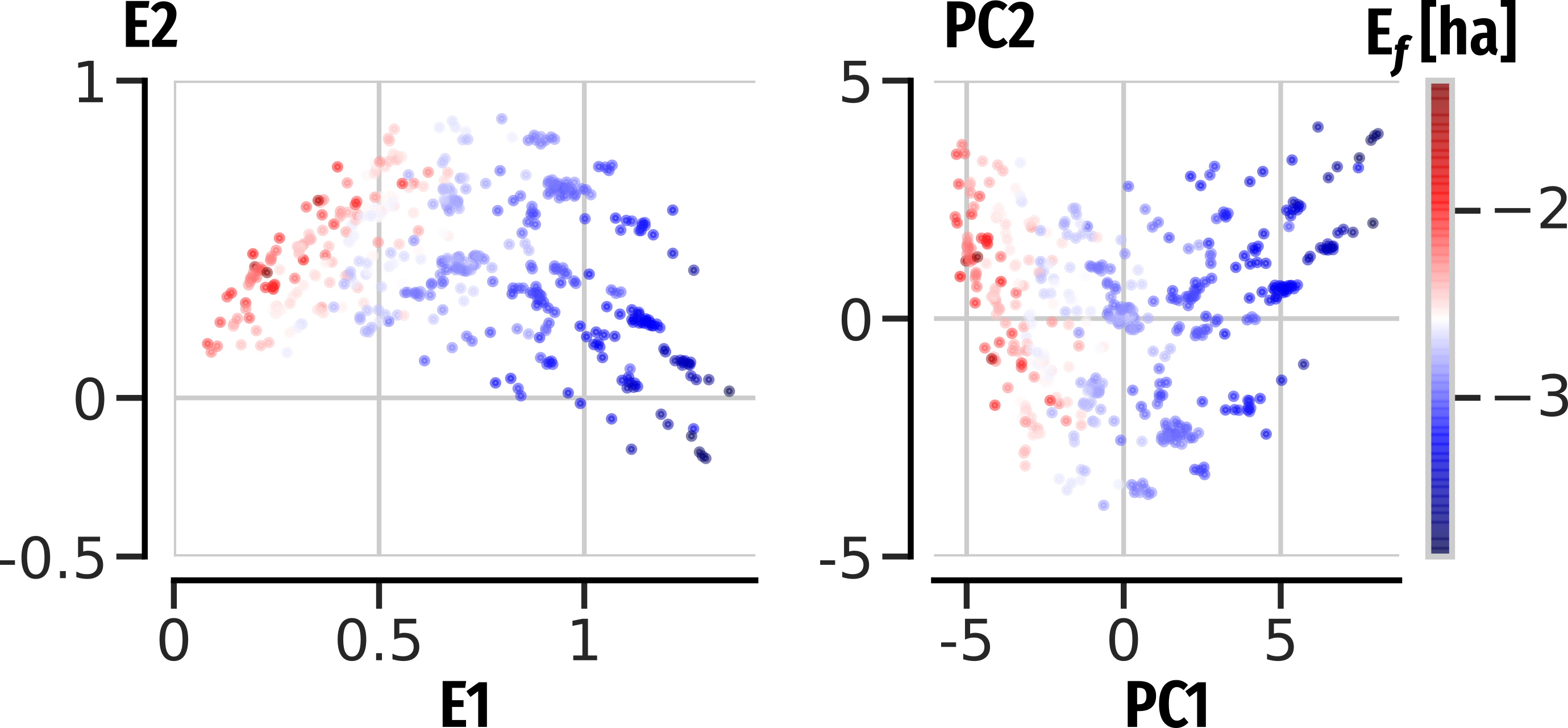} 
          \caption{Two-dimensional encrypted neural network (NN) auto-encoding with components $\text{E}1,~\text{E}2$ of Bag-of-Bonds\cite{bob} representation vectors for the first 20000 QM9\cite{qm9} test molecules in (a). In (b) we show the first two principal components $\text{PC}1,~\text{PC}2$ resulting from principal component analysis using the scikit-learn python package\cite{sklearn}. The color coding shows corresponding formation energies $E_{\text{f}}$.
          }
     \label{fig:qml_net_dimred} 
 \end{figure*}

 \begin{figure}[htb]
          \centering           \includegraphics[width=0.35\textwidth]{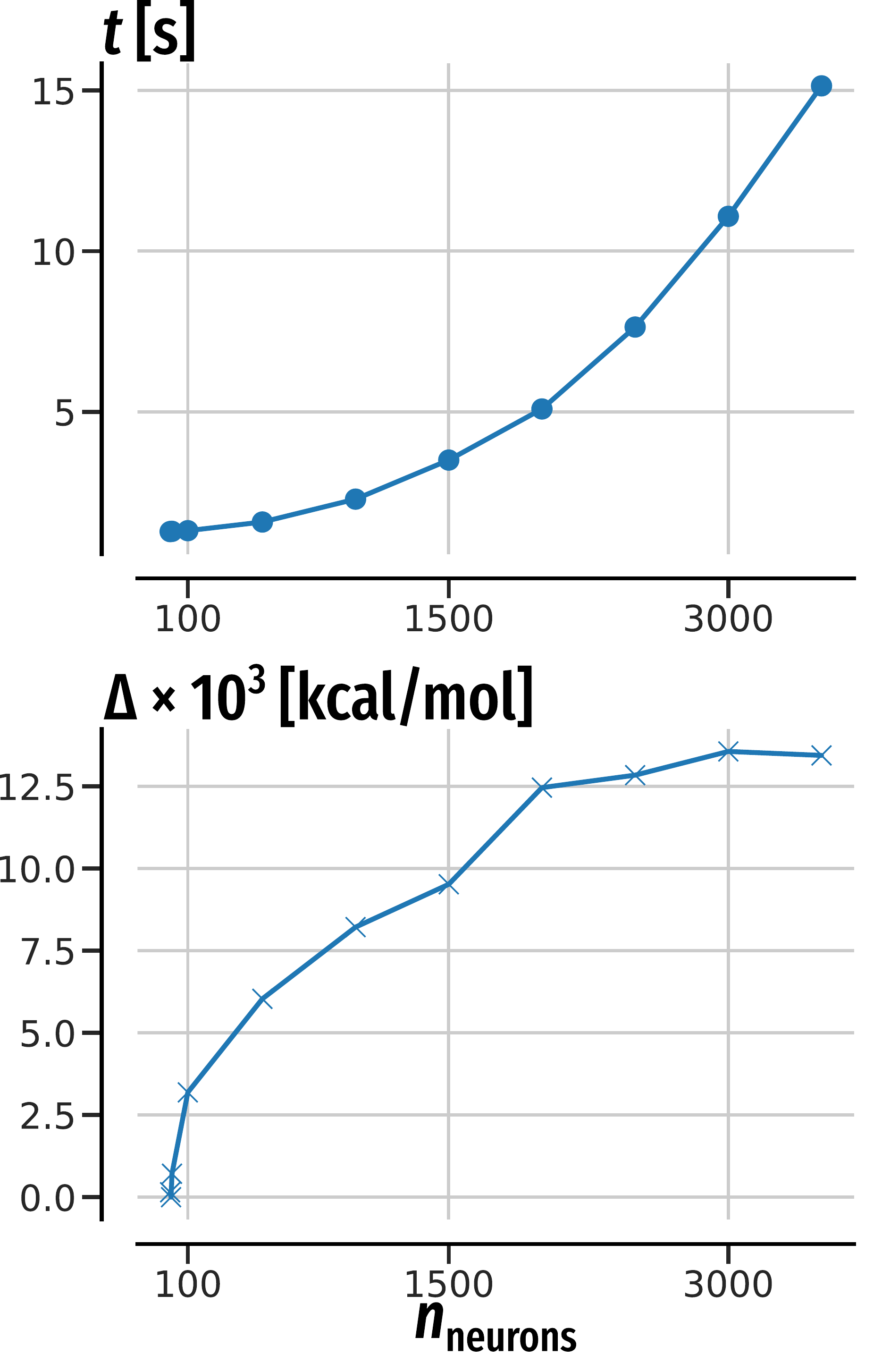} 
          \caption{Average per molecule encrypted prediction time for different numbers of neurons $n_{\text{neurons}}$ used in the neural network as well as numerical error associated with the encrypted predictions.
          }
     \label{fig:qml_net} 
 \end{figure}

\begin{figure*}[htb]
          \centering           
    \includegraphics[width=0.6\textwidth]{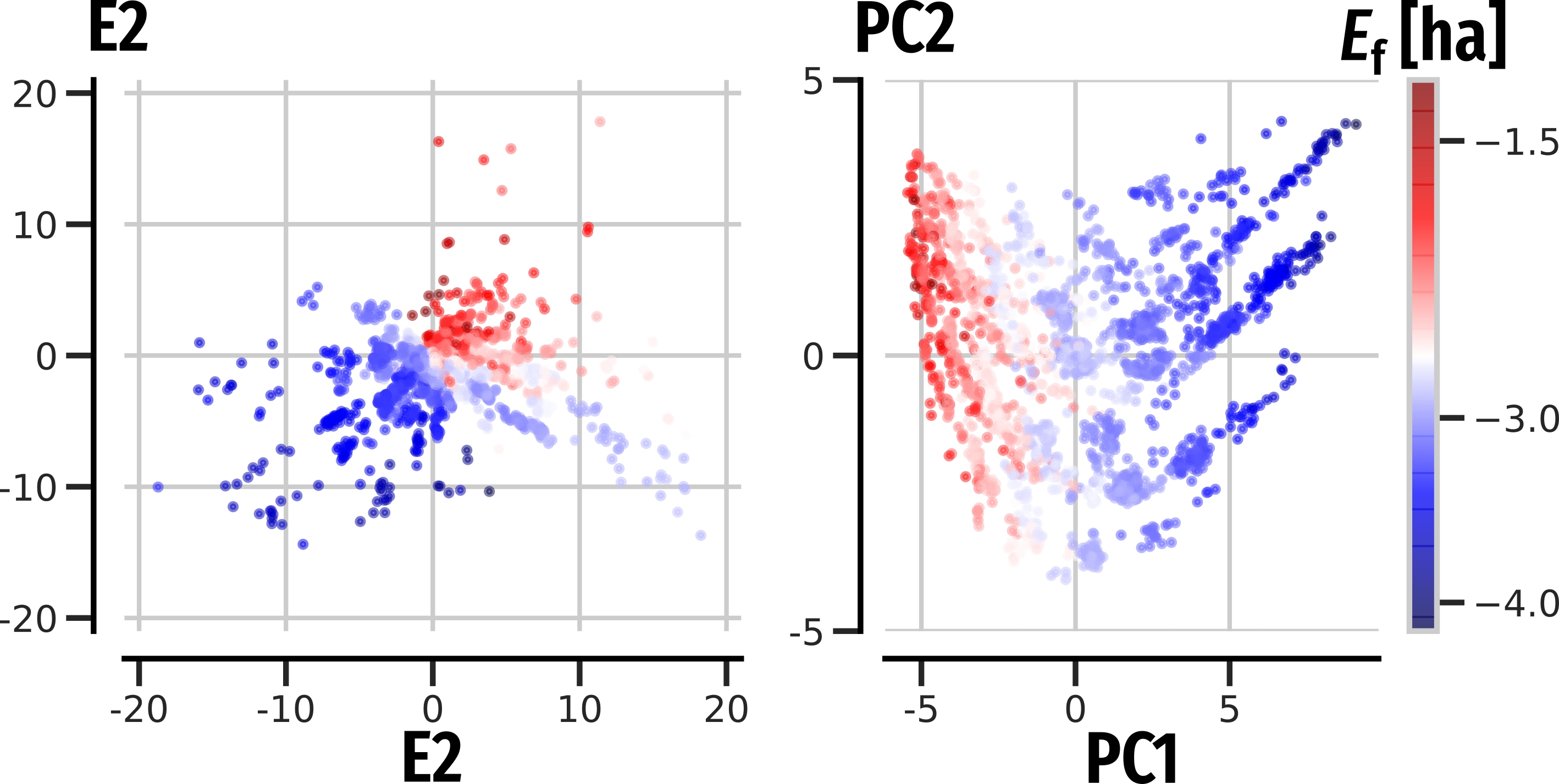} 
          \caption{
          Two-dimensional encrypted neural network (NN) symmetric stacked autoencoder auto-encoding with components $\text{E}1,~\text{E}2$ of Bag-of-Bonds\cite{bob} representation vectors for a random subset of 20000 QM9\cite{qm9} molecules in (a). In (b) we show the first two principal components $\text{PC}1,~\text{PC}2$ resulting from a plaintext principal component analysis. The color coding shows corresponding formation energies $E_{\text{f}}$.
          }
     \label{fig:qml_net_dimred} 
 \end{figure*}

 \begin{figure}[htb]
          \centering                \includegraphics[width=0.35\textwidth]{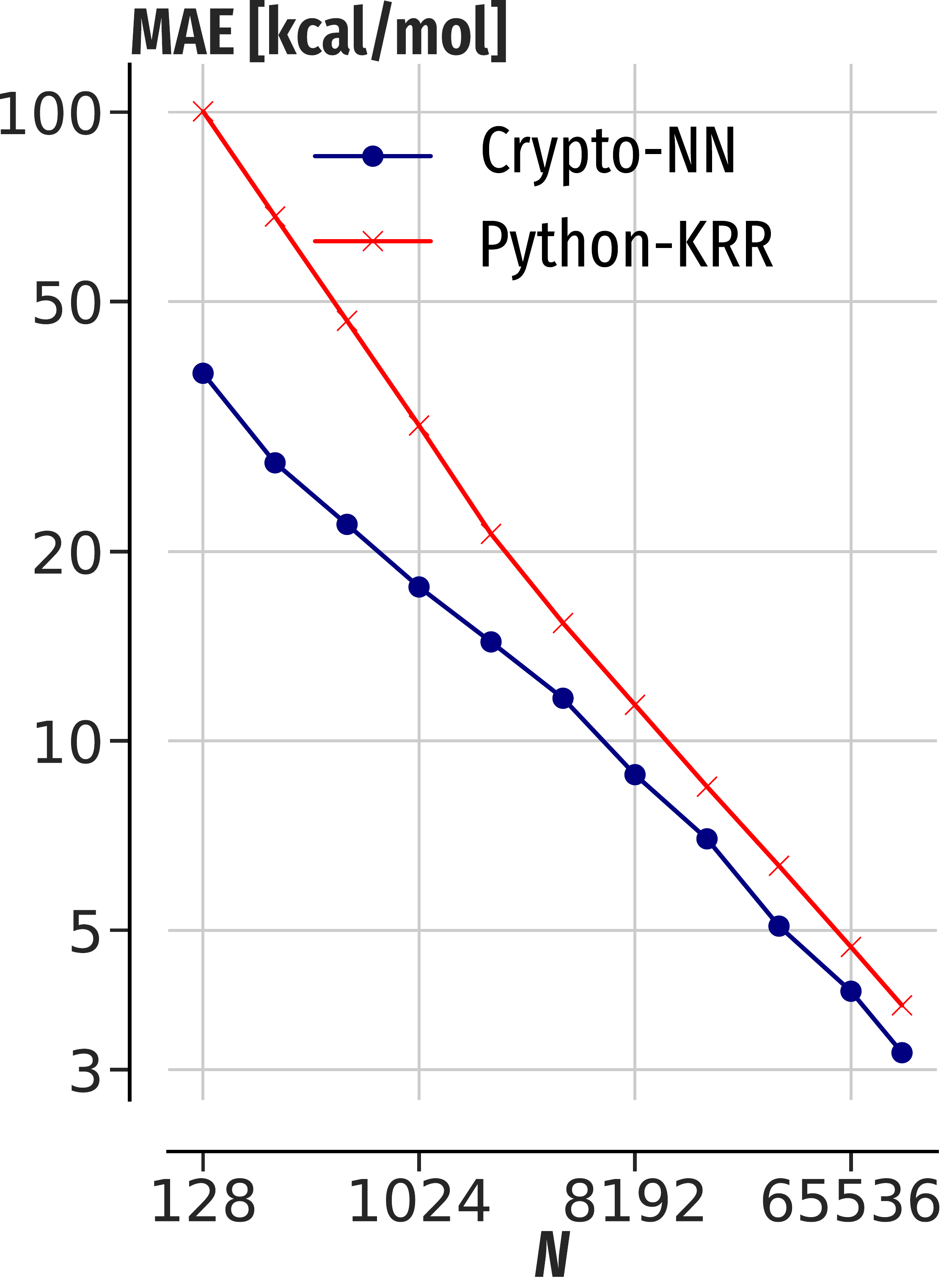}
          \caption{Mean absolute error (MAE) of QM9\cite{qm9} atomization energies as a function of the number of training molecules $N$ for encrypted neural network predictions as well as corresponding MAE of plaintext kernel ridge regression predictions with python using a Gaussian kernel.
          }
     \label{fig:qml_net} 
 \end{figure}

We consider encrypted predictions in an \textit{honest-but-courious} security scenario with encrypted neural networks. In contrast to the MASCOT\cite{mascot} kernel ridge regression implementation, here data privacy is only guaranteed if both parties strictly follow the appointed protocol $f$. We use a different oblivious transfer protocol based on CrypTen\cite{crypten2020} and PyTorch\cite{NEURIPS2019_9015}. As before we test the two-party case with \textcolor{SkyBlue}{\textbf{A}} and \textcolor{LimeGreen}{\textbf{B}}. The workflow is the same as for the encrypted kernel ridge regression discussed in the manuscript. The main difference is that the target function $f$ is predicted with a neural network. A neural network consists of a large number of so-called artificial neurons inspired by biological neural networks\cite{geron}. Neurons are assembled in layers and their connections are called edges forwarding information from the input to the output layer. The weights determine the signal strength between neurons. Training of a neural network reduces the error of neural network predictions of the input w.r.t. to training labels. In this process called backpropagation, the change in average error is traced back to the individual weights of the neural network. Subsequently, the weights are adjusted to reduce the overall error in an iterative process. The architecture of the neural network consists of eight fully connected rectified linear unit layers with $n_{\text{neurons}} = 250$ neurons each and one linear output layer. Each feature of the BoB\cite{bob} molecule representation vector corresponds to one input neuron. Training is performed with unencrypted calculations. Subsequently, the neural network and query molecules are encrypted and can be used to predict the encrypted query molecules.
\\
Due to the higher computational efficiency of this implementation, we calculated a learning curve for the complete QM9\cite{qm9} dataset. We find excellent agreement between the learning curves resulting from the python implementation and encrypted predictions (s. Fig.~\ref{fig:qml_net}). In this case, the neural network slightly outperforms the learning curve of an unencrypted kernel ridge regression model with a Gaussian kernel function. Each encrypted prediction takes about 1.3 seconds when using $n_{\text{neurons}} = 250$ neurons per layer. Testing the prediction time as a function of the number of neurons for a fixed number of layers reveals (s. SI Fig.~5a) quadratic scaling with a favorable scaling constant: A prediction with $n_{\text{neurons}} = 3000$ neurons per layer (eight layers total) takes 10.4 seconds.

\subsubsection{Secret data comparison}
\label{sec:pca}

\textcolor{SkyBlue}{\textbf{A}} and \textcolor{LimeGreen}{\textbf{B}} want to measure the chemical overlap between their respective secret data domains without revealing individual data points. While this might reveal some information about the chemical domain of interest, the degree of information leakage revealed by the overlap can be controlled by the number of function evaluations that both parties have to agree on.
\\
To implement the secret data comparison, we use an encrypted encoder.
Autoencoders are artificial neural networks that learn dense encodings of input data and do not require data labels. Encodings usually have a much smaller dimension than the original input dimension, making autoencoders well-suited for dimensionality reduction. An encoder is made up of layers that map to latent space and a decoder network that decodes the latent representation of the input data. The output dimension of the decoder is identical to the input dimension of the encoder. Training the autoencoder aims to reduce the decoder reconstruction error. An encoder with a single two-dimensional linear layer and the mean-squared error as a loss function will qualitatively resemble the first two components of principal component analysis\cite{geron} maximizing the variance of the encoding in two dimensions. To implement the encrypted encoder, we use the same \textit{honest-but-curious} multi-party computation framework for neural networks based on CrypTen\cite{crypten2020} as before.
\\
Apart from data comparison, another possible application of such an encoder would be to reduce the dimension of the previously studied representations, leading to higher numerical stability and faster predictions by encoding into two dimensions. Since the encoder-decoder networks' goal is to maximize the variance in two dimensions, the two-dimensional encoding (s. Fig.~\ref{fig:qml_net_dimred}a) results in a distribution of points similar to that resulting from principal component decomposition (s. Fig.~\ref{fig:qml_net_dimred}b).

\newpage
\newpage

\bibliography{literature}
\end{document}